\documentclass[12pt]{iopart}
\usepackage{epsfig}
\usepackage{graphicx}
\usepackage{citesort}

\def\be{\begin{equation}}
\def\ee{\end{equation}}
\def\bea{\begin{eqnarray}}
\def\eea{\end{eqnarray}}

\begin{document}
\title[Review of Bose-Einstein or HBT correlations ...]
{{\bf  Review of HBT or Bose-Einstein correlations in high energy heavy ion collisions} }
\author{T. Cs\"org\H{o}}
\address{MTA KFKI RMKI, H - 1525 Budapest 114, P.O.Box 49, Hungary}

\begin{abstract}
A brief review is given on the discovery and the first five decades of the
Hanbury Brown  - Twiss effect and its generalized applications in high energy nuclear
and particle physics, that includes a meta-review.
Interesting and inspiring new directions are also highlighted, 
including for example source imaging, lepton and photon interferometry, 
non-Gaussian shape analysis as well as many other new directions. 
Existing models are compared to two-particle correlation measurements 
and the so-called RHIC HBT puzzle is resolved.
Evidence for a (directional) Hubble flow is presented and the conclusion is confirmed 
by a successful description of the pseudorapidity dependence
of the elliptic flow  as measured in Au+Au collisions by the PHOBOS Collaboration. 
\end{abstract}



\medskip
\begin{flushleft}
    {\small\it 
Child, how happy you are sitting in the dust, 
    playing with a broken twig all the morning.\\
I smile at your play 
with that little bit of a broken twig. \\
I am busy with my accounts,
adding up figures by the hour. \\
Perhaps you glance at me and think,
"What a stupid game to spoil your morning with!"\\
Child, I have forgotten 
the art of being absorbed in sticks and mud-pies.\\
I seek out costly playthings, 
and gather lumps of gold and silver. \\
With whatever you find you create your glad games,\\ 
I spend both my time and my strength over things I never can obtain. \\
In my frail canoe I struggle to cross the sea of desire, 
	and forget that I too am playing a game.  \\
	(R. Tagore: Playthings)}
\end{flushleft}
\medskip

\section{Introduction}
First I attempted to give a brief summary of the numerous achievements of particle interferometry 
from 1955 to 2005, to celebrate the Golden Jubilee: the 50 year anniversary of 
the Hanbury Brown - Twiss effect.  As it is clearly impossible to summarize 
50 years of efforts and discoveries of the order of one thousand scientific papers in a few pages, 
I ended up in creating a meta-review.
I had to be very brief in reviewing - see my talk in ref.~\cite{india-talk05} for more details,
historical remarks and animated illustrations. 

\section{The Hanbury Brown - Twiss effect}
One of the most inspiring books that I have read recently was R. Hanbury Brown's 
brief autobiography -- Boffin. I recommend the reading of this book for all students
all who enter the field of particle interferometry 
in high energy particle and nuclear physics~\cite{Boffin}.

As this note is contributed to the ICPA-QGP 2005 conference organized in India,
I should mention that  Robert Hanbury Brown was born here 
in Aruvankadu, Nilgiri Hills, India in 1916 -- 
he passed away in Andover, England, in 2002.  Similarly to him Richard Q. Twiss 
was also born in India  and they both had loved this country.

Hanbury Brown had been trained as an engineer and he was marked with an 
ingenious talent as well as a subtle sense of humor. 
Let me quote his book on how he reflects on his most famous discovery:
``{\it I was a long way from being able to calculate, whether it would be sensitive enough to
measure a star. To do that one has to be familiar with photons and as an engineer my 
education in physics had stopped far short of the quantum theory. Perhaps just as well, 
otherwise like most physicists I would have come to the conclusion that the thing would not 
work -- ignorance is sometimes a bliss in science." ...
``In fact to a surprising number of people the idea that the arrival of photons at 
two separated detectors can ever be correlated was not only heretical but patently absurd,
and they told us so in no uncertain terms, in person, by letter, in print, 
and by publishing the results of laboratory experiments, which 
claimed to show that we were wrong ... " } . 
Opponents cited also from sacred books of Quantum Mechanics, for example quoting 
Dirac~\cite{Dirac}: {\it ``Interference between two different photons can never occur." }
Vigorous objections were based on two laboratory experiments, the first of which
was performed by the group of J\'anossy at KFKI, my home institute in Budapest, Hungary.
Their paper concluded~\cite{Janossy}, that {\it `` ...  in agreement with quantum theory,
    the photons of two coherent light beams are independent from each other ."}
However, Hanbury Brown and Twiss analyzed this and other experiments, 
and pointed out ~\cite{HBT-brief} that to observe the HBT effect one needs a 
very intensive source of light with a very narrow bandwidth, 
such as an isotope lamp that neither of these counter-experiments
had.
		
The first stellar intensity interferometer, 
the pilot model was set up at Jodrell Bank in England 50 years ago, in 1955. 
This equipment was used to make the first measurement of the angular diameter of a 
main sequence star, Sirius ~\cite{HanburyBrown:1956pf}.
This pilot experiment demonstrated the value of the method of intensity correlations 
and was the precursor to the construction of the Stellar Intensity Interferometry in Narrabi,
Australia,  completed in 1963. The giant reflectors of the Narrabi Observatory had
parabolic surfaces of 22 feet diameter and a focal length of 36 feet, 
mounted on a circular railway truck, that had 618 feet diameter,
or 73 \% of the 843 feet diameter of the AGS accelerator 
of Brookhaven National Laboratory in Upton, LI, USA.

During the last 50 years, the study of intensity correlations became
a broad field not only in quantum optics, solid state physics and
astronomy, but extended to include the investigations of 
correlations among various type and number of particles in high energy particle and nuclear physics,
\cite{NA22.multi,UA1.multi,NA44.kaons,STAR-firstHBT,PHENIX-HBTbigkt,L3-twopi,L3-3pi,STAR-nonid,STAR-3pi}.

\section{Particle interferometry from 1955 to 2005 - a review of reviews}
The field of particle interferometry
or femtoscopy in high energy particle and heavy ion physics is firmly established:
there is even  a book written about the topic~\cite{Weiner-book}.
A collection of the early articles on Bose-Einstein correlations in high
energy physics is also available~\cite{Weiner-collection}.  
Here I collected some of the most inspiring review papers of the field.

For newcomers, I recommend to start with the summary of 
L\"{o}rstad from 1989~\cite{bengt-rev}, written 
at a period when preliminary correlation data indicated signatures for a first order
QGP to hadron transitions in O + Au collisions at CERN SPS energies.
Boal, Gelbke and Jennings~\cite{boal-g-j-rev} summarized boson and fermion intensity
correlations in subatomic physics, with emphasis on final state interactions.
The review of  Bauer, Gelbke and Pratt~\cite{bauer-gelbke-pratt-rev} 
is recommended for techniques applied in low and intermediate energy heavy ion collisions
in particular for the noting of the energy dependence of the effective source sizes
in nucleon-nucleon intensity correlations. A pedagogical work aimed at the 
identification of experimental difficulties in two-boson interferometry was
compiled by Zajc~\cite{bill-rev}, a must for every pedestrian entering this field.
Schutz~\cite{schutz-rev} summarized hard photon interferometry in heavy ion physics.
Harris  and M\"{u}ller overviewed the suggested signatures for
the production of a quark-gluon plasma, 
including Bose-Einstein correlations~\cite{harris-muller-rev}. 
Light particle correlation data in heavy ion collisions 
at intermediate and low energies was compiled by Ardouin~\cite{ardouin-rev}.  
Data and theory of correlations in relativistic heavy ion collisions were
summarized by Heinz and Jacak~\cite{heinz-jacak-rev}.  
The multivariate Gaussian approach to the parameterization of particle interferometry
in relativistic heavy ion collisions was highlighted by  
Wiedemann and Heinz~\cite{wiedemann-heinz-rev}.
The invariant Buda-Lund particle interferometry method
was summarized by me and  L\"orstad~\cite{cst-bengt-rev}.
The review of Weiner~\cite{weiner-rev} highlighted the applications of quantum
optical methods in boson interferometry in high energy particle and nuclear physics.
Bose-Einstein correlations in electron-positron annihilations at LEP were 
introduced and reviewed by Kittel~\cite{kittel-rev2},
summarized at the $Z$ fragmentation~\cite{kittel-rev3},
and at the $W^+ W^-$ decays~\cite{kittel-rev1}.
Model independent methods, similarities between interferometry results in
electron-positron annihilation data and in relativistic heavy ion collision
data, critical review of the resonance and Coulomb effects, 
limitations of the Gaussian approach and the Bertsch-Pratt and Yano-Koonin
parameterizations were discussed~\cite{cs-rev}
together with applications of source image reconstruction and quantum
optical methods like squeezed states and pion lasers.
Tomasik and Wiedemann reviewed central and non-central 
particle interferometry data after the first results at RHIC became
accessible~\cite{tomasik-wiedemann-rev}.
Alexander~\cite{alexander-rev} reviewed Bose-Einstein and Fermi-Dirac
correlations in particle physics, with focus on the mass and transverse
mass dependence of the radius parameters of these correlations in 
electron-positron annihilation at LEP. 
Heinz~\cite{heinz-rev} reviewed the concepts applied
in high energy heavy ion physics, including particle interferometry.
The review of  
Padula~\cite{sandra-rev} starts with an excellent 
historical overview of interferometry in the 1960's and 1970's,
covering the current formalism and squeezing of mass-modified bosons and fermions.
These reviews well summarize the first 50 years of particle interferometry.

\section{Interesting new directions}

{\it Improvement on the two-particle Coulomb corrections}.
If $\lambda < 1$, either partial coherence develops in the source, or some fraction
of the pions comes from the extended, 
halo part of the source~\cite{Grassberger:1976au,cs-rev}.  This influences,
how the relative separations of the particle pairs are distributed,
hence the weight assigned to various parts of the two-particle Coulomb
wave function. CERES~\cite{Adamova:2002wi}
was the first experiment to apply this averaging self-consistently. By now,
STAR~\cite{Adams:2003ra},
PHENIX~\cite{PHENIX-HBTbigkt} and PHOBOS~\cite{Back:2004ug} have also utilized
this method suggested by Bowler~\cite{Bowler:1991vx}, Sinyukov and collaborators~\cite{Sinyukov:1998fc}.

{\it Many-body Coulomb effects for finite sources}.
The core-halo type of problem exists not only in the construction of the Bose-Einstein correlation
function of two charged particles, but in the determination of the
Bose-Einstein correlation function of three or more charged particles.
The many-body Coulomb corrections are notoriously difficult,
but the circumstances are favorable
in high energy heavy ion and particle physics for the application of an approximate,
cluster expansion method, that is based on an asymptotically correct form of the 
multiparticle Coulomb wave function~\cite{Alt:1999cs}.

{\it Intensity correlations of non-identical particle pairs}
are new and promising tools in heavy ion physics, 
given the large possible combinations of various type of particles.
This method can not only be used to learn about the temporal sequence of particle
emission, but also to learn about the strong final state interactions between various
type of particle pairs~\cite{Voloshin:1997jh,STAR-nonid}.
	 
{\it Intensity correlations of penetrating probes} provides information not only on the
phase-space distribution of the decoupling system at the time of freeze-out
but also adds the information about the temporal evolution of the fireball.

{\it a) Photon interferometry} attracted a strong theoretical activity
in order to utilize correlations of direct photons to search for quark gluon plasma phase 
suggested first by Makhlin~\cite{Makhlin:1989bg}.
This field is pioneered by Srivastava and collaborators,
\cite{Bass:2004de,Srivastava:1993hw,Srivastava:1993pt,Srivastava:1993js}
as well as by the organizers of this conference~\cite{Alam:2003gx}.
Earlier papers highlighted to photon interferometry at the CERN SPS energy domain
\cite{Pisut:1995vr,Timmermann:1994kb,Ostendorf:1992ah}, where 
the first experimental results on photon interferometry were reported recently
by WA98 in Pb+Pb collisions~\cite{Aggarwal:2003zy}.
				     
{\it b) Lepton interferometry} is also  based on  
penetrating probes that can escape from the hadronic fireball
during its time evolution, hence lepton correlations can  give
information on the volumes of homogeneity  at any stage of the expansion,
at least in principle. The first case study has just been published 
by Alam, Mohanty and collaborators~\cite{Alam:2004eu} using 
a 3+1 dimensional, relativistic hydrodynamical model
with spin dependent invariant amplitudes and a bag model equation of state.

The key experimental question for photon and lepton interferometry  is if one can find 
a transverse momentum window where photons or leptons  from a quark gluon
plasma phase dominate the single particle spectrum, overcoming 
contributions e.g. gammas from the $\pi^0$ decays, or in case of electrons, 
from Dalitz and open charm decays and Drell-Yan processes.
				     
{\it Q-boson interferometry} has been investigated theoretically by Zhang, Padula, Anchiskin
and collaborators in refs. ~\cite{Zhang:2002wv,Anchishkin:1999vg,Anchishkin:2000wk}.
It remains to be seen if q-boson correlations can ever be observed in experiments of 
high energy physics.

{\it Similarities between Bose-Einstein correlations in different reactions}
might indicate an important limitation of our understanding of the 
particle correlations in $e^+ + e^-$, hadron + hadron, d+A and A+B reactions. For example,
the mass and the transverse mass dependencies of the Bertsch - Pratt radius
parameters in these reactions seems to be very similar, but the scales  and
the reaction mechanisms are completely different~\cite{kittel-rev1,kittel-rev2,kittel-rev3,alexander-rev,cs-rev}.

{\it Continuous emission and escaping probabilities} has to be taken into account,
if particle emission is not confined to a narrow freeze-out hypersurface, but is
continuous during the time evolution of the system.  This scenario has been considered
by Grassi, Hama and collaborators, who developed first 
the continuous emission model~\cite{Grassi:2000ke}.
This model was further improved with the help of
the introduction of the escaping probabilities and a self-consistent theoretical
formulation of the particle emission from hydrodynamical sources during the
whole period of expansion~\cite{Sinyukov:2002if}.
	 
{\it Initial conditions that fluctuate event by event} are seen 
in realistic Monte-Carlo calculations.  Averaging over the final states 
of hydrodynamics started from fluctuating initial
conditions yields HBT radii closer to the measured values
as compared to the case when the initial conditions
are first averaged over and the hydrodynamic evolution starts from the smoothed
initial conditions, as shown with the help of a 3+1 dimensional
relativistic hydrodynamical  codes developed by the SPHERIO
collaboration~\cite{Socolowski:2004hw}.

{\it Rapidity dependence } of the HBT radii was explored by
PHOBOS at RHIC and NA49 at CERN SPS.
For the side, out and longitudinal components only weak rapidity 
dependence was found in both cases.  The Yano-Koonin-Podgoretskii 
fit revealed that the longitudinal flow profile 
is nearly boost invariant in both reactions~\cite{Back:2004ug,Kniege:2004pt}.

{\it Azimuthally sensitive HBT measurements} 
push the multi-variate Gaussian parameterization to its limits. 
The transverse momentum and the azimuthal angle 
dependence of the 3 by 3 symmetric radius matrix provide important constraints for the
dynamics of the effective source in these collisions, pioneered by M. Lisa
and collaborators~\cite{Lisa:2000ip}. STAR data on asHBT indicate~\cite{Adams:2003ra}, 
that the effective source is extended in the direction perpendicular to
the impact parameter, similarly to the geometry of the nuclear overlap.

{\it Beyond the Gaussian approximation:}
{\it a) The Edgeworth and Laguerre expansions} 
utilize complete sets of polynomials that are orthogonal 
with respect to given weight functions.
The method was worked out by S. Hegyi and the
present author~\cite{Csorgo:2000pf}.
The Edgeworth expansion has been applied to
quantify the non-Gaussian features of Bose-Einstein correlations by the L3 and
the STAR collaborations~\cite{Achard:2002ve,L3-3pi,Adams:2004yc}.

{\it b) The L\'evy index of stability}, $0 < \alpha \le 2 $
is a new parameter of the two- and three-particle
Bose-Einstein correlation functions~\cite{Csorgo:2004ch,Csorgo:2003uv,Csorgo:2004sr}.
It characterizes the power-law tails of 
L\'evy stable distributions, which  appear in physical systems 
where the final distribution is obtained as convolution
of many elementary random steps. 
If the distribution functions of the elementary
steps are characterized by finite means and variances, central limit theorems
determine that the limit distribution has to be a Gaussian and $\alpha = 2$. 
Generalized central limit theorems state that for certain elementary
processes with infinite variance or infinite mean a limiting distribution
exist, but the limit distribution develops a power-law tail and $\alpha < 2$ .

{\it c) Imaging methods} were developed by 
Brown and Danielewicz\cite{Brown:1997ku} to reconstruct the relative coordinate 
distribution of the source by inverting the two-particle correlation function with a kernel,
based on known final state interaction and symmetrization effects. This method was
applied first at the AGS energies~\cite{Panitkin:2001qb}. 
Preliminary PHENIX results indicate~\cite{Chung:2005ra}
the existence of a long tail in the relative coordinate distributions 
in d+Au and Au+Au collisions at the maximal RHIC energies.

{\it Multi-boson symmetrization effects and pion lasers}
are explored with the pion-laser model of S. Pratt~\cite{Pratt:1993uy}, 
that I could understand only from ref.~\cite{Chao:1994fq}, which lead to the  
analytical solution of the model with Zim\'anyi~\cite{Csorgo:1997us,Zimanyi:1997ur}.
Among others we found that in the rare gas limit two and three-particle symmetrization can be
performed perturbatively, within Poisson distributed clusters. 
More theoretical and experimental work is needed to investigate these effects
under conditions of RHIC and LHC reactions.

{\it In-medium hadronic mass modifications} can signal partial
$U_A(1)$ symmetry restoration, measurable with two-pion Bose-Einstein 
correlations~\cite{Vance:1998wd}. Current STAR Au+Au HBT radii were shown
in agreement with a model assuming chiral symmetry restoration~\cite{Cramer:2004ih}.
The quantum freeze-out  problem of in-medium mass shifted bosons
has been solved~\cite{Asakawa:1998cx} correcting the 
idea of Andreev and Weiner~\cite{Andreev:1995cb}. It was
extended to the case of fermions~\cite{Panda:2000wr}.
A liquid of mass-modified hadrons is thus signalled by  
back-to-back correlations between the detected particle - antiparticle pairs. 
 
Let me add the following important comment here.
The most recent studies of the phase structure of strongly interacting matter
with the help of  lattice QCD calculations suggest that the transition from
the hadronic phase to a this form of matter, the quark-gluon plasma, 
happens at a critical temperature of $T_c = 164 \pm 2$ MeV and the form of
the transition is a cross-over, where hadrons might exist above the critical
temperature in a broad range of baryo-chemical potentials up to $\mu_B \approx 350$
MeV, where the transition changes to a second order phase transition.
At even larger baryochemical potentials, a first order phase transition is seen,
and the critical temperature decreases slightly with increasing $\mu_B$-s,
see ref.~\cite{Fodor:2004nz} for further details.
If indeed the degrees of freedom in a fluid are dominated by non-hadronic states,
and if hadronization and freeze-out happens simultaneously in a time-like
deflagration as suggested in ref.~\cite{csorgo-csernai}, 
then the back-to-back particle-antiparticle 
correlations have to vanish.  Hence the excitation function of the
these correlations changes drastically at the transition from a hadronic to a
non-hadronic liquid. Hence it is sensitive  to the creation of a quark gluon plasma,
which underlines the importance of search for squeezed states in ultra-relativistic
heavy ion collisions.

    For advanced studies, I recommend the following works and important contributions 
to particle interferometry in high energy physics from its first decade,
~\cite{Goldhaber:1959ex}-\cite{Glauber:1963tx}
its second~\cite{Kopylov:1974th}-\cite{Yano:1978gk},
third~\cite{Podgoretsky:1982xu}-\cite{Gyulassy:1989pp},
fourth~\cite{Baechler:1991pz}-\cite{Bearden:1998aq}
and fifth decade~\cite{Gyulassy:2001zv}-\cite{Lisa:2003ze}.

\section{The ``RHIC HBT puzzle" and its resolution}

Before contrasting the experimental results at RHIC with theoretical expectations
and fits, let me go back first to the theory of particle interferometry
for expanding sources formulated by Pratt~\cite{Pratt:1984su,Pratt:1986cc}.
The directional dependence of the HBT correlation function was investigated 
by Hama and Padula as early as in 1987~\cite{Hama:1987xv},
who observed that the longitudinal effective radius parameter is proportional
to the proper-time of freeze-out, and proportional to the square root of  temperature over energy.
Their eq. (7)  of ref.~\cite{Hama:1987xv} 
can be considered as one of the first precursors to the studies of the ratios or 
the differences between $R_{out}$ and $R_{side}$.
Makhlin and  Sinyukov derived a famous and experimentally well tested formula for
boost-invariant particle emitting sources: $R_{long} = \tau_f \sqrt{T_f/m_t}$,
ref.~\cite{Makhlin:1987gm}.  Bertsch introduced~\cite{Bertsch:1989vn} 
the naming convention (out,side,longitudinal) for the components of the relative momentum  
that are parallel with the mean transverse momentum, 
perpendicular to the longitudinal and the mean momentum, 
and parallel with the longitudinal direction, respectively, with an influential 
graphical illustration, supported by Monte-Carlo cascade simulations by Bertsch, Gong 
and Tohyama~\cite{Bertsch:1988db}.
Using a Gaussian approximation, the popular fit parameters thus became 
$\lambda$, $R_{out}$, $R_{side}$, and $R_{long}$. 

\subsection{HBT result at CERN SPS energies}
The first preliminary results from the NA35 experiment
in O+Pb reactions at CERN SPS energies indicated a signal for the first order QGP-hadron transition,
$R_{out}$ was found to be much larger, than $R_{side}$, 
a three sigma effect~\cite{Humanic:1988ny,Bamberger:1988kd}.
These data were  described by Padula and Gyulassy 
in terms of a quark gluon plasma model but also in terms of a
conventional hadronic resonance gas model~\cite{Padula:1988uf,Gyulassy:1988yr}.
Kaon interferometry has been suggested to
increase the selectivity of these correlation measurements~\cite{Gyulassy:1989pp}.

The preliminary NA35 data were changed and the difference between
the out and the side HBT radius parameters in S+Pb collisions was found to  vanish 
within errors~\cite{Baechler:1991pz,Alber:1995dc}.
This observation was published by the second generation experiment NA44,
which attempted to determine precisely all the three radius components
at mid-rapidity in a broad transverse mass interval~\cite{Beker:1994qv}, valid not
only for pions but also for kaons~\cite{Beker:1994qs,Bearden:1999ix}.
The results indicated an approximate equality and simultaneous scaling
of the radius parameters of the correlation functions,
namely that these radius parameters within rather small experimental
errors turned out to be rather similar, obeying a common dependence
on the transverse mass of the pair. At the same time the NA44 collaboration
observed the scaling of the single particle spectra, indicating that the
slope parameter increases with increasing transverse mass of the particles~\cite{Bearden:1996dd}.
Both of these effects were explained in terms of a three-dimensionally expanding,
axially symmetric, suddenly decaying fireball
~\cite{Csorgo:1995bi,Csorgo:1995vf}.
The existence of  scaling limiting cases in certain domain of the parameter space
was a focal point for 
the formulation and subsequent development of this Buda-Lund hydro model.

\subsection{HBT predictions for RHIC energies}
Soon after the observation of the approximate equality and transverse mass scaling
of the radius parameters of the Bose-Einstein correlation function (HBT radii for short),
Gyulassy and Rischke predicted that the HBT radius 
in the out direction will exceed the HBT radius in the side direction,
predicting $R_{out}/R_{side} > 1$, values  reaching up to 4 or even 20 in case of an
ideal first order QGP to hadron phase transition,
\cite{Rischke:1995cm,Rischke:1996em}.
These detailed predictions about a slowly burning log of QGP 
obtained a lot of experimental attention, in contrast to the earlier and less well-known but 
in fact more successful model of ref.~\cite{csorgo-csernai}, that predicted
$R_{out}\approx R_{side} \approx R_{long}$ in a broad transverse mass interval at RHIC
for a QGP which has reduced entropy density and harmonizes suddenly 
at every location from a supercooled state.
Presently neither of the four experimental  predictions made in
ref.~\cite{csorgo-csernai} are in disagreement with the 
known features of particle production in Au+Au collisions
at RHIC as far as I know, hence the picture of a suddenly frozen, supercooled QGP
has not yet been invalidated by experimental observations at RHIC.

\subsection{Less Unpromising Models and the Au+Au HBT data at RHIC}
The first detailed measurements of Bose-Einstein or HBT correlations by STAR at RHIC
~\cite{STAR-firstHBT} have excluded the validity of the slowly burning quark-gluon plasma picture
predicted in ref.~\cite{Rischke:1995cm,Rischke:1996em}. These measurements were confirmed
by PHENIX~\cite{PHENIX-HBTbigkt} and the approximate equality of the HBT radii was
established in a large transverse momentum interval. Recently, PHOBOS has extended
these measurements to a broad rapidity interval without changing the 
overall picture~\cite{Holzman:2004pb}.

Of the order of 50 models were shown to be unable to describe these observables
at RHIC. The difficulty faced here by some of the theoretical approaches to
describe the data is frequently referred to as the ``RHIC HBT puzzle"~\cite{Gyulassy:2001zv}.

With the help of my students, M. Csan\'ad and A. Ster, we have scanned the literature
and attempted to determine the list of less unpromising models, that cannot be
excluded with the tools of mathematical statistics, using HBT data at RHIC.
The following models were found to pass this HBT test:
\begin{itemize}
\item
The multi-phase transport model (AMPT) of Lin, Ko and Pal~\cite{Ko:2002iz}.
\item
The hadronic cascade model calculation of Humanic~\cite{Humanic:2002iw}.
\item
The Buda-Lund hydro model~\cite{Csorgo:2002ry,Csanad:2003sz,Csanad:2004cj,Csanad:2004mm}.
\item
The Cracow single freeze-out thermal model~\cite{Broniowski:2002wp}.
\item
The Blast-wave model as implemented by Retiere and Lisa~\cite{Retiere:2003qb,Retiere:2003kf}.
\item
The parameterized time-dependent expanding souce model of Renk~\cite{Renk:2003gn,Renk:2004yv}.
\item
Ref.~\cite{Cramer:2004ih} that investigates the onset of the chiral phase 
transition and its effects on the HBT radii. Effectively, this model also leads  
to a Buda-Lund type of parameterization.
\end{itemize}
As the number of models that pass the HBT test is fairly sizeable at present,
additional criteria are needed to select the most relevant models. 

\section{Evidence for a Hubble flow in Au+Au at RHIC}
To constrain the class of the not obviously wrong models, additional observables
are needed. Let me chose here the identified single particle spectra
of PHENIX, the pseudorapidity distribution measured by BRAHMS and PHOBOS,
and the transverse mass dependence of the HBT radius parameters as measured
by PHENIX and STAR. Humanic's cascade, the blastwave model, the Buda-Lund hydro model
and the Cracow hydro model pass these additional criteria as well.
In my talk I have illustrated~\cite{india-talk05}, how a transverse and a longitudinal
Hubble constant can be extracted from the blastwave, the Buda-Lund and the Cracow model
fits to these data. Within errors, all of these three models are agreement
with an approximately direction independent Hubble type of flow profile,
where the transverse and the longitudinal Hubble constants are within errors
the same, $u_{long} = H_{long} r_{long}$ and  $u_{transv} = H_{transv} r_{transv}$, with
$H_{long}=H_{transv} = 0.13  \pm 0.02$ $fm^{-1}$.
This result is well illustrated by the upper plots in Fig.1.

These observations are confirmed by a perfect description of the rapidity dependence
of the elliptic flow with the help of the ellipsoidal generalization of the
Buda-Lund hydro model, as illustrated on the lower plots of Figure 1. 
    These data by the PHOBOS collaboration were not
reproduced with the help of hydrodynamical models earlier. The lower right panel
of Figure 1 illustrates, that the PHOBOS data are not only perfectly described
with the ellipsoidally symmetric generalization of the Buda-Lund hydro model,
    but also that the PHOBOS data can be scaled to the theoretically predicted
    form of the scaling function. 

    This ellipsoidally symmetric Buda-Lund model 
is based on direction dependent Hubble constants~\cite{Csanad:2003qa}.
The asymptotic emergence of a direction independent Hubble flow is seen analytically
in the new families of exact analytic solutions of 
non-relativistic hydrodynamics that are the basis this Buda-Lund
hydro model~\cite{Csorgo:1998yk,Csorgo:2002kt,Csorgo:2001xm}.
\begin{figure}[ht]
\begin{center}
\vspace{-0.5cm}
\begin{center}
	\includegraphics[width=3.0in]{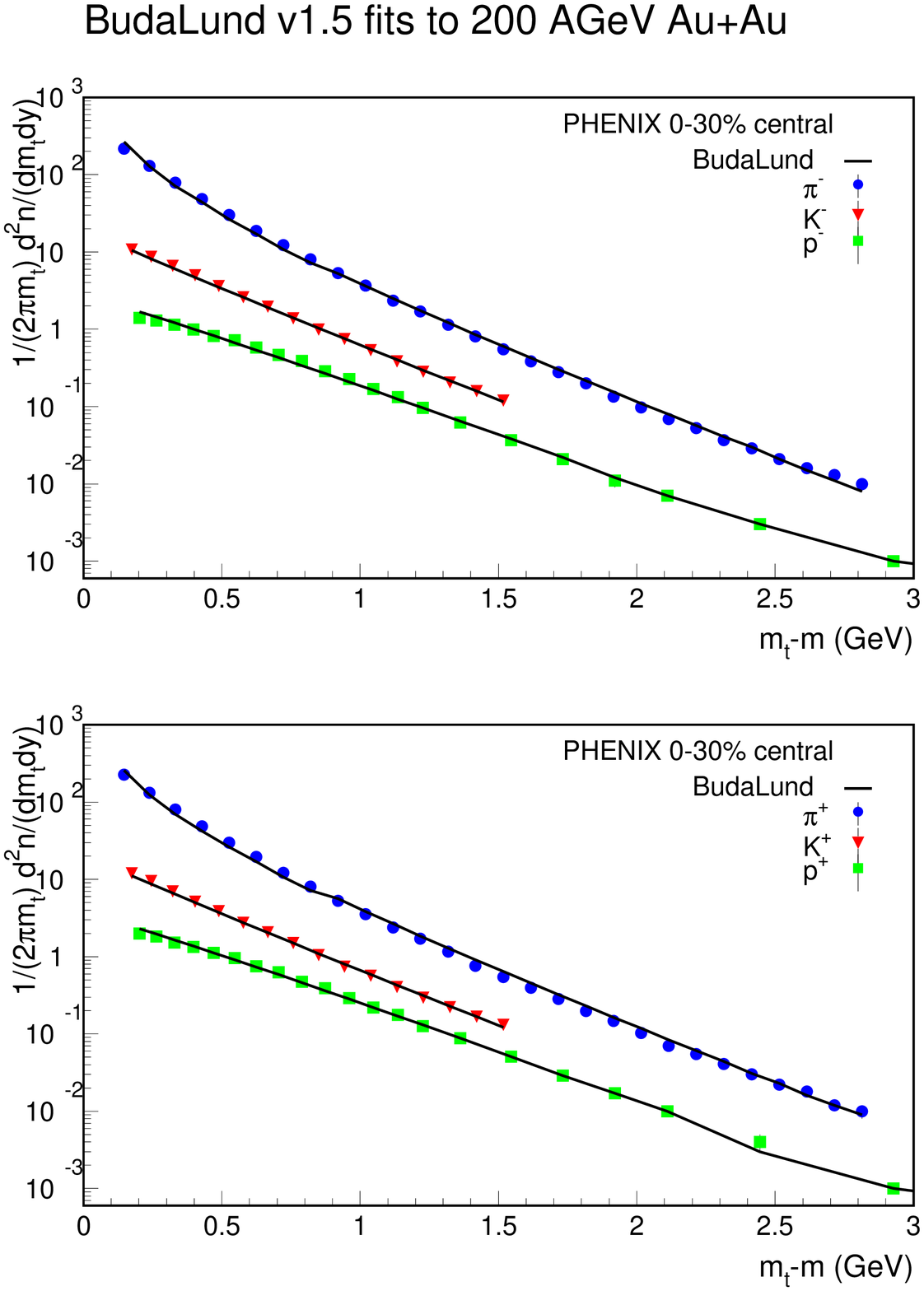}
	\includegraphics[width=3.0in]{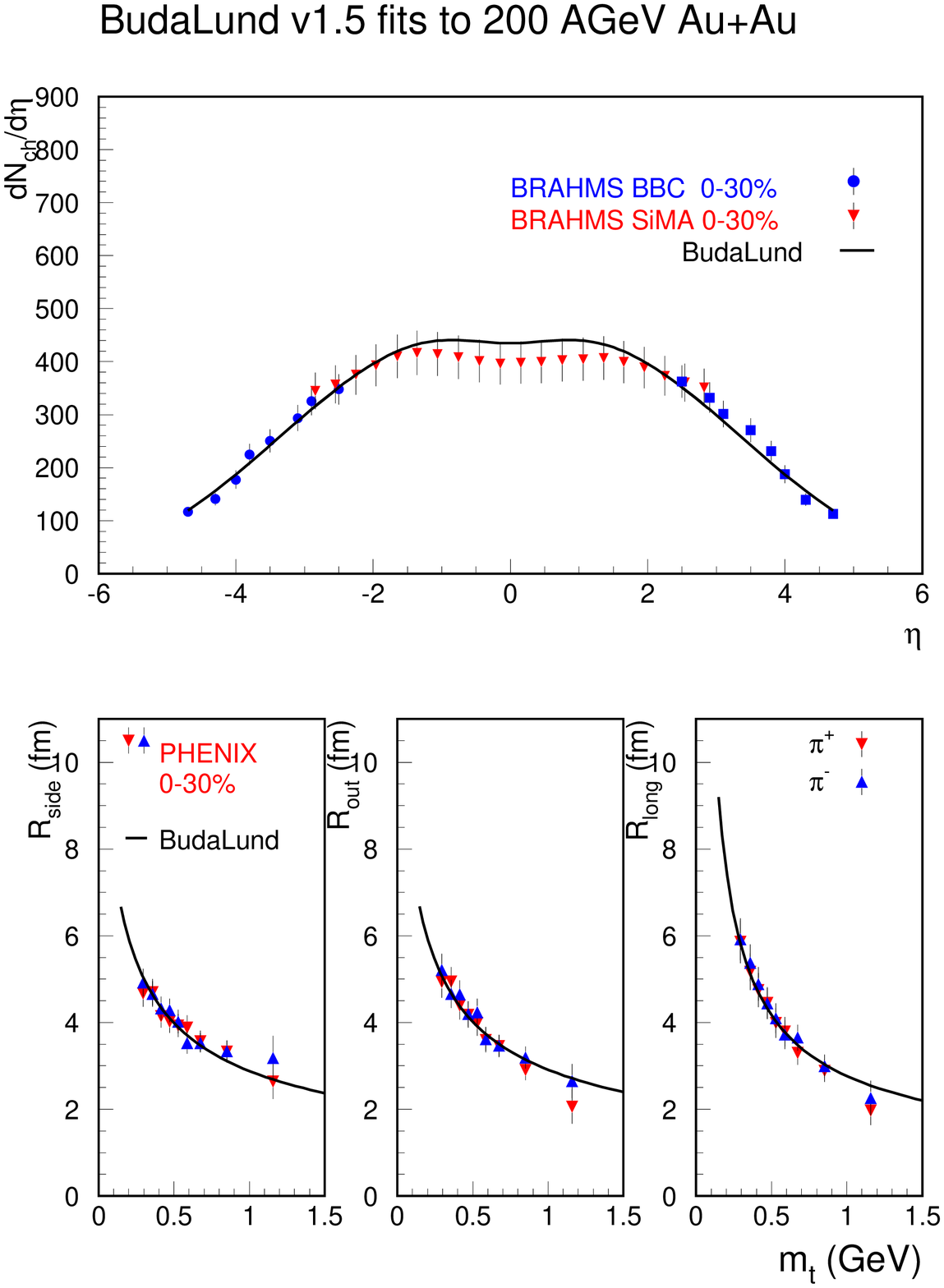}\\
\includegraphics[width = 2.8in]{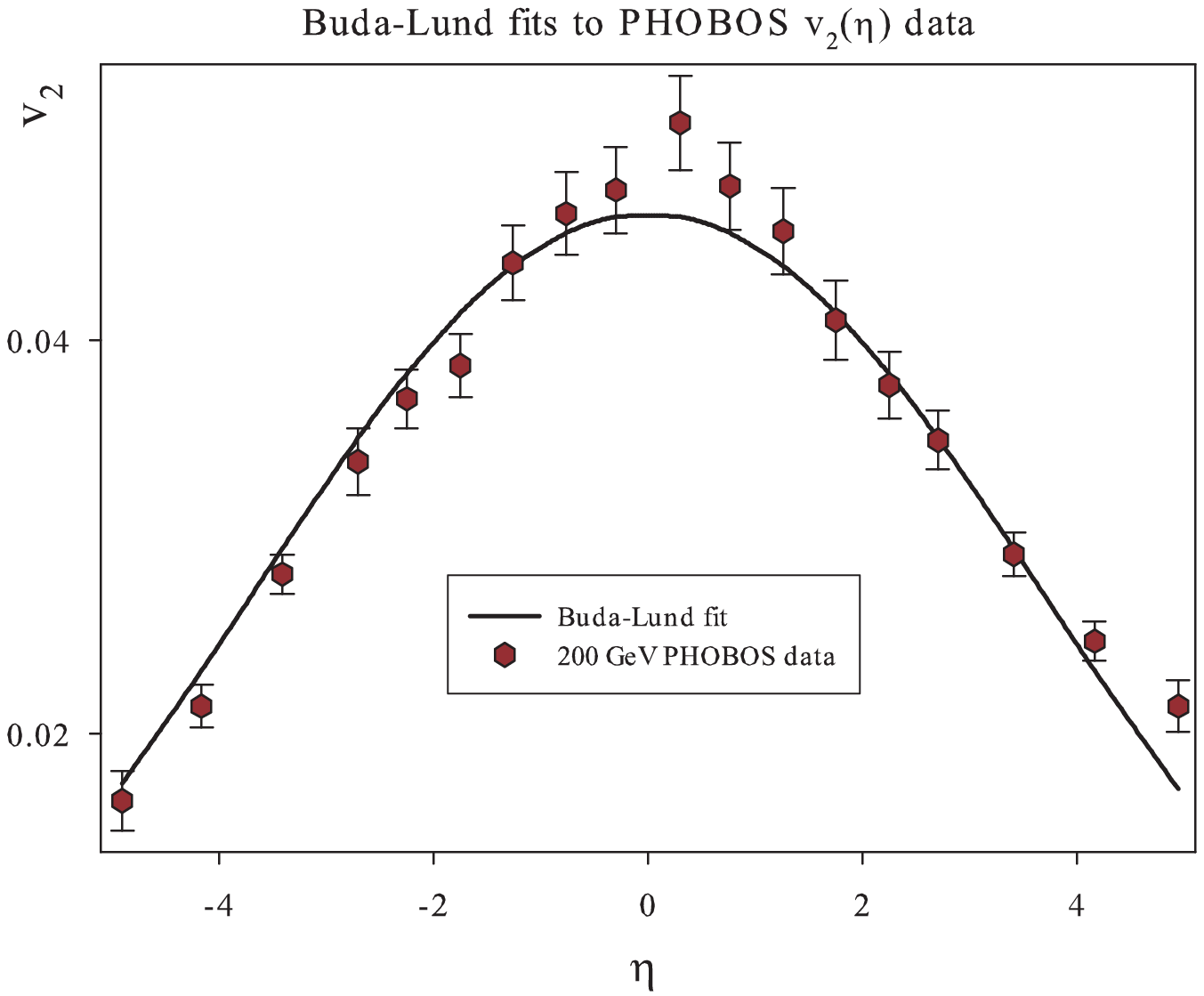}
\includegraphics[width = 3.0in]{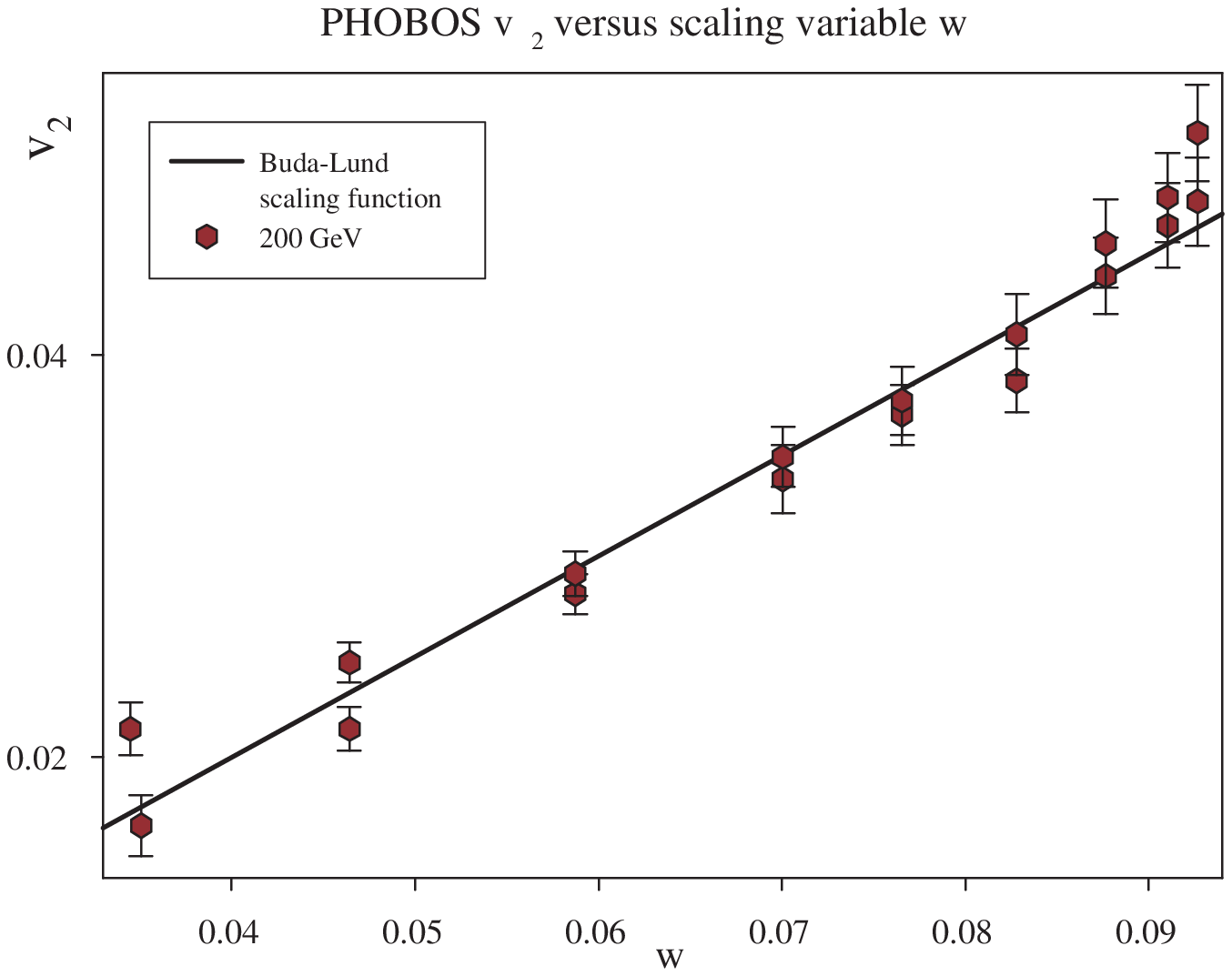}
\end{center}
\caption[*]{
\label{fig:130-spectra}{\small The upper  panels show a  simultaneous Buda-Lund fit  to final
Au+Au data at $\sqrt{s_{NN}} = 200$ GeV from 
refs.~\cite{Bearden:2001qq},~\cite{Adler:2003cb,PHENIX-HBTbigkt}. 
The fit parameters are summarized in Table 1 of ref.\cite{Csanad:2004mm}.
The ellipsoidally symmetric Buda-Lund hydro model is fitted~\cite{Csanad-prep}  
to preliminary PHOBOS  $v_2(\eta)$ data at 200 GeV on the lower left panel~\cite{PHOBOS-v2-qm02}.
In the lower right panel, these PHOBOS data on $v_2(\eta)$ are re-scaled and compared 
to the scaling function $v_2(w) = I_1(w)/I_0(w)$ predicted 
by the Buda-Lund hydro model~\cite{Csanad:2003qa}.
}}
\end{center}
\end{figure}

\section{Conclusions}

After a historic introduction and a review of reviews I have highlighted some of the
interesting new directions, including interferometry with penetrating probes like
photons and leptons and the quantification of the non-Gaussian behaviour of the correlation
functions. It may well be that a non-trivial energy dependence of these correlations
reveals itself not in the scale parameters like the HBT radii but  in the shape 
parameters like the L\'evy index of stability $\alpha$. Indeed, for  p+p reactions,
this parameter has been linked~\cite{Csorgo:2004sr} to the anomalous dimension of QCD which follows
a non-trivial energy dependence due to the running of the strong coupling constant.
I also noted that the disappearance of the particle-antiparticle back-to-back 
correlations at a certain colliding energy can be a new signal for a transition 
from a hadronic fluid to a non-hadronic fluid,
hence to the onset of the deconfinement transition. More experimental efforts are
needed to look for squeezed states and chiral symmetry restoration at the RHIC energies.

The HBT data at RHIC provide important and selective constraints for the model
builders and indicate that a large number of models implemented the space-time
evolution of the source incorrectly. Many models failed to describe these observables,
but this is not a puzzle. Mysteries surround the HBT effect from the very beginning of
its discovery.  I noted, that there was a successful prediction~\cite{csorgo-csernai} 
in 1994 for the simultaneous
equality of the HBT radii in Au+Au reactions  at RHIC  in a broad transverse mass interval, and 
all the qualitative predictions of this model are in agreement with the
present data. Hence currently it cannot be excluded that a suddenly hadronizing
supercooled QGP is present in these reactions.  I presented
the list of the less unpromising models - theoretical
descriptions that pass the HBT test in Au+Au collisions at RHIC. At present
7 such models were found in the literature.
Then I suggested to try to determine the common part of these models and to increase
the sensitivity of the test by comparing these models to the single particle spectra,
the HBT radii and to the elliptic flow measurements at RHIC. 
I have argued that the hadronic cascade model of Humanic, the Buda-Lund hydro model,
the blastwave model and the Cracow model passes all these tests. 
In an attempt to determine the common features of these
models, the transverse and the longitudinal Hubble constants were found to be
the same within errors in the best fits to the data by the Buda-Lund, the Blastwave
and the Cracow models, providing an evidence for a nearly fully developed Hubble flow
and an indication of deconfinement temperatures reached 
in central Au+Au collisions at RHIC~\cite{Csanad:2004mm,Csanad:2005qh}.
These observations are confirmed by a perfect description of the rapidity dependence
of the elliptic flow with the help of the Buda-Lund hydro model. 
Based on a forthcoming manuscript~\cite{Csanad-prep} 
I have shown that the PHOBOS rapidity dependent
elliptic flow data, previously resisting a description in a hydrodynamic picture,
actually agree with the theoretically predicted scaling function if the proper
scaling variable $w$ is determined. The solution of the HBT puzzle at RHIC
thus naturally provides a solution to the puzzling rapidity dependence of the
elliptic flow at RHIC as well.

\section*{Acknowledgments:}
I would like to express my sincere thanks to the Organizers of this meeting for creating 
an excellent scientific atmosphere for ICPA-QGP 2005. 
This work was supported by  following grants:
OTKA T034269, T038406, T049466, the OTKA-MTA-NSF grant INT0089462, the
NATO PST.CLG.980086 grant and the exchange program
of the Hungarian and the Polish Academy of Sciences.

\end{document}